\documentclass[final,5p,twocolumn]{elsarticle}

\usepackage{graphicx,epsfig}

\usepackage{amssymb}
\usepackage{amsthm}

\journal{Physics Letters A}

\begin{document}

\begin{frontmatter}

%% Title, authors and addresses

%% use the tnoteref command within \title for footnotes;
%% use the tnotetext command for theassociated footnote;
%% use the fnref command within \author or \address for footnotes;
%% use the fntext command for theassociated footnote;
%% use the corref command within \author for corresponding author footnotes;
%% use the cortext command for theassociated footnote;
%% use the ead command for the email address,
%% and the form \ead[url] for the home page:
%% \title{Title\tnoteref{label1}}
%% \tnotetext[label1]{}
%% \author{Name\corref{cor1}\fnref{label2}}
%% \ead{email address}
%% \ead[url]{home page}
%% \fntext[label2]{}
%% \cortext[cor1]{}
%% \address{Address\fnref{label3}}
%% \fntext[label3]{}

\title{Salecker-Wigner-Peres clock and average tunneling times}

%% use optional labels to link authors explicitly to addresses:
%% \author[label1,label2]{}
%% \address[label1]{}
%% \address[label2]{}

\author[uepg]{Jos\'{e} T. Lunardi}
\ead{jttlunardi@uepg.br}

\author[cord]{Luiz A. Manzoni\corref{ca}}
\cortext[ca]{Corresponding author. Tel.: +1(218) 299 3395; fax: +1(218) 299 4308}
\ead{manzoni@cord.edu}

\author[cord]{Andrew T. Nystrom}
\ead{atnystro@cord.edu}

\address[uepg]{Departamento de Matem\'atica e Estat\'{\i}stica,
Universidade Estadual de Ponta Grossa\\ Av. General Carlos
Cavalcanti, 4748. Cep 84030-000, Ponta Grossa, PR, Brazil}

\address[cord]{Department of Physics, Concordia
College, 901 8th St. S., Moorhead, MN 56562, USA}

\begin{abstract}
The quantum clock of Salecker-Wigner-Peres is used, by performing a post-selection of the final state, to obtain average transmission and reflection times associated to the scattering of localized wave packets by static potentials in one dimension. The behavior of these average times is studied for a gaussian wave packet, centered around a tunneling wave number, incident on a rectangular barrier and, in particular, on a double delta barrier potential. The regime of opaque barriers is investigated and the results show that the average transmission time does not saturate, showing no evidence of the Hartman effect (or its generalized version).
\end{abstract}

\begin{keyword}
%% keywords here, in the form: keyword \sep keyword
Salecker-Wigner-Peres quantum clock\sep time dependent tunneling\sep average tunneling times\sep Hartman effect
%% PACS codes here, in the form: \PACS code \sep code
%\PACS 03.65.Xp\sep 73.40.Gk\sep 03.65.Ta
%% MSC codes here, in the form: \MSC code \sep code
%% or \MSC[2008] code \sep code (2000 is the default)
\end{keyword}

\end{frontmatter}

%% \linenumbers

%% main text
\section{Introduction}

As it is well known, the Hamiltonian of a quantum mechanical system must be bounded from below due to stability requirements and, as a consequence, there is no self-adjoint time operator canonically conjugated to the Hamiltonian \cite{CNi68} (see, however, \cite{Olk09}), hence posing serious difficulties to define a meaningful time scale for quantum phenomena such as tunneling.

Several approaches have been considered to circumvent the above difficulty, leading to many definitions of quantum tunneling time, among which the dwell time and the phase time are the most prominent (see \cite{HSt89, LMa94, Win06-2, Win06-1} and references therein). However, the time scales proposed until now are generally plagued by difficulties which prevent their interpretation as a traversal time in the general case.  For instance, a serious controversy in the interpretation of many such time scales is related to the Hartman effect, which asserts that in the stationary case the time saturates for a static opaque potential \cite{Har62} (also see \cite{ORS02, LLB02} for a generalized version).

However, one cannot properly address the problem of quantum tunneling time in the stationary case because the very concept of a particle having a definite tunneling time requires that it be represented by a \textit{localized} wave packet (see, for example, \cite{FHa88}). With this in mind, in this work we reexamine the tunneling of a wave packet through a static potential by using the quantum clock formalism introduced by Salecker and Wigner \cite{SWi58} and further refined by Peres \cite{Per80}. Our main concern here is to scrutinize whether the Hartman effect, both in its original and generalized versions, emerges or not for a wave packet.

The Salecker-Wigner-Peres (SWP) clock \cite{SWi58,Per80} (see also \cite{Lea93,CLM09}) consists of a quantum rotor weakly coupled to the system (the tunneling particle), and it can be shown to be equivalent to the Larmor clock \cite{FHa88, But83, LAe89}, prepared and observed in a particular set of states, in the limit of large spin -- for a detailed treatment see \cite{SCo93}. In this framework a ``time operator" (not canonically conjugated to the Hamiltonian) can be defined \cite{Per80}, and the application of such an operator to quantum tunneling was extensively analyzed by Leavens and McKinnon \cite{LMc94}, who found that its expectation value only produces a meaningful result after a calibration procedure is adopted for the clock readings. Recently, Park \cite{Par09} considered the one-dimensional scattering of a wave packet by a rectangular potential barrier and found that the expectation value of that ``time operator" in the asymptotic final states gives the dwell time averaged with respect to the components of the initial wave packet, which, however, cannot be considered a traversal time because it does not distinguish between transmitted and reflected particles \cite{Par09}.

In this paper, instead of associating the clock readings to expectation values of such a ``time operator", we follow Peres' approach in the study of the time of flight \cite{Per80}. According to this method the transmission (reflection) time, which is obtained by following the peak of the clock's wave function, is given by the time derivative of the phase shift of the transmission (reflection) amplitude with respect to the perturbation potential. We argue that in order to address the question of the time spent in the barrier region by a particle that is \textit{eventually transmitted} (\textit{reflected}) one \textit{must} perform a \emph{post-selection} of the final state. Then, making use of an \textit{asymptotic condition} ensuring that for asymptotic times the incident and the transmitted/reflected wave packets are well localized far from the potential, it is possible to trace out the particle's degrees of freedom in order to ``read" the clock's final state. The average times obtained through this procedure are shown to have sensible physical properties in the context of quantum tunneling of gaussian wave packets through static symmetric potentials.

This work is organized as follows. In section \ref{clock} we briefly review the formalism introduced by Peres \cite{Per80} and its generalization to the time dependent situation (also see \cite{Lea93, Par09}), introducing the necessary steps to take into account only particles that are eventually transmitted (reflected) by the potential. We discuss some properties of the average times obtained and compare them with other time scales suggested in the literature. In Section \ref{tun} we study the behavior of these average times for the scattering of a gaussian wave packet, centered around a tunneling wave number, by a single rectangular barrier and a double delta barrier, with emphasis on the opaque regime and the possible emergence of the Hartman effect and, in particular, its generalized version. In section \ref{concl} we present our conclusions.

%%%%%%%%%%%%%%%%%%%%%%%%%%%%%%%%%%%%%%%%%%%%%%%%%%%%%%%%%%%%%%%%%%%%%%%%%%%%%%%%%%%%%%%%%%%%%%%%%%%%%%%%%%%%%%%%%%%%%%%%%%%%%%%%%%%

\section{The Salecker-Wigner-Peres Clock}
\label{clock}

The Salecker-Wigner clock \cite{SWi58}, as considered by Peres \cite{Per80}, is just a quantum rotor with Hamiltonian given by
\begin{equation}
H_c = -i \hbar \omega\frac{\partial}{\partial\theta}\; ,
\end{equation}
with $\omega \equiv 2\pi/(N\tau)$, where $\tau$ is the time resolution of the clock and $N= 2j+1$ is the dimension of the Hilbert space (with $j$ being a non-negative integer or half-integer).

The eigenfunctions $u_m(\theta)= \frac{1}{\sqrt{2\pi}}e^{i m\theta}$ ($0\leq \theta < 2\pi$) of $H_c$ have eigenvalues $\mathcal{V}_m = m\hbar\omega$
with $m= -j, \ldots , +j$. A more convenient orthogonal basis for the clock is given by $v_r(\theta)$ ($r=0, \ldots , N-1$) \cite{Per80}
\begin{equation}
v_r(\theta) =  \frac{1}{\sqrt{N}}\sum_{m=-j}^j \exp \left(- 2\pi i\frac{r m}{N} \right) u_m(\theta)\; ,
\end{equation}
which has the advantage of having a sharp peak at $\theta = 2\pi r/N$ for large $N$, so that the clock's pointer indicates the $r$th ``hour" (with uncertainty $\pm \frac{\pi}{N}$). Under time evolution such states change as $v_r(\theta)\to v_r(\theta-\omega t)$, thus making possible to associate a time duration $t$ to an arbitrary interval just by observing the translation of the peak of a free clock's wave function $v_r$ in that interval.

We are interested in measuring the time associated with the one-dimensional scattering of a particle of mass $\mu$ by a (real) static potential $V(z)$ confined to the region $0\leq z\leq a$. Following \cite{Per80}, the particle and the clock are coupled
through the interaction Hamiltonian
\begin{equation}
H_I = \mathcal{P}(z)H_c\, ,
\end{equation}
where the projection operator  $\mathcal{P}(z)$ satisfies  $\mathcal{P}(z)\!=\!1$ if $z\!\in \![0,a]$ and equals zero otherwise. The clock is assumed to
be initially at the state $v_0(\theta)$. Thus, the general form of the initial wave function of the whole system is
\begin{equation}
 \Psi_{inc}(\theta,z,t) = \Phi (z,t)v_0(\theta)\, ,
\end{equation}
where $\Phi (z,t)$ is a wave packet corresponding to a particle incident from the
left of the potential which, from the symmetries of the problem \cite{Mer}, can be written as
\begin{equation}
\Phi (z, t) = \int \frac{dk}{2\pi} \;A(k) e^{i\left(kz - \frac{E(k)t}{\hbar}\right)}\, ,
\label{wavepack}
\end{equation}
with all integrals in $k$ going from $-\infty$ to $+\infty$. The Fourier coefficient $A(k)$ is assumed to be centered around a positive $k_0$ corresponding to a tunneling energy.

After the scattering the particle and the clock become entangled due to the interaction $H_I$ and the general form of the wave function is \cite{Lea93}
\begin{equation}\label{coupled}
\Psi(\theta,z,t) = \frac{1}{\sqrt{N}}\sum_{m=-j}^j \Phi^{(m)}(z,t)u_m(\theta)\, ,
\end{equation}
where
\begin{equation}
\Phi^{(m)} (z, t) = \int \frac{dk}{2\pi} \; A(k) \psi^{(m)}
 (k,z) e^{-i\frac{E(k)t}{\hbar}}\, .
\label{wavepackm}
\end{equation}

The wave function $\Psi (\theta,z,t)$ satisfies the Schr\"{o}dinger equation with the Hamiltonian $H=\frac{p^2}{2\mu}+V(z)+H_I$. The orthogonality of the $u_m(\theta)$'s, together with the fact that they are eigenfunctions of $H_c$, imply that
$\psi^{(m)}(k,z)$ satisfy the time-independent Schr\"{o}dinger equation for energy $E(k)$ and potential
\begin{equation}
V^{(m)}(z) = V(z) + \mathcal{V}_m \mathcal{P}(z)\, ,
\end{equation}
where $\mathcal{V}_m = m\hbar\omega $. In order to ensure a weak coupling between the clock and the system it is assumed that $\mathcal{V}_m$ is a small perturbation, i.e.,
$|E|\gg\mathcal{V}_m$ and $|V(z)-E|\gg\mathcal{V}_m$.

Since we are interested in measuring the transmission time asymptotically after the interaction, let us focus on the transmitted wave packet.
The solution of the above stationary problem in the region to the right of the potential $V^{(m)}(z)$ is
$$
\psi_{trans}^{(m)}(k,z) = T^{(m)}(k)e^{ikz}
= \left| T^{(m)}(k) \right| e^{i\varphi_T^{(m)}(k)+ikz}\, .
$$
By using the weak coupling condition we can assume that
$\left| T^{(m)}(k) \right| \simeq \left| T(k) \right|$ and expand the phase $\varphi_T^m(k)$ up to first order in $\mathcal{V}_m$ \cite{Per80}:
\begin{equation}\label{1order}
\varphi_T^{(m)}(k) \simeq \varphi_T (k) + \mathcal{V}_m  \left( \frac{\partial }{\partial\mathcal{V}_m}\varphi_T^{(m)}(k)\right)_{\mathcal{V}_m=0}\, ,
\end{equation}
where $\varphi_T (k)$ denotes the phase of the transmission coefficient corresponding to the unperturbed potential $V(z)$. Finally, substituting the above expressions in (\ref{coupled})-(\ref{wavepackm}), and using the explicit form of the functions $u_m(\theta)$, the asymptotic transmitted wave function is given by
\begin{eqnarray}
\Psi_{trans}(\theta, z,t) &\!=\!& \int \frac{dk}{2\pi} \; A(k) T(k) e^{i(kz-\frac{E(k)t}{\hbar})} \nonumber \\
&&\,\,\,\,\,\,\,\,\,\,\,\,\,\,\,\,\times\, v_0\left(\theta - \omega t_c^T(k) \right)\, ,
\label{psi_tr}
\end{eqnarray}
where
\begin{equation}\label{tc}
t_c^T (k)= - \hbar \left( \frac{\partial }{\partial\mathcal{V}_m}\varphi_T^{(m)}(k)\right)_{\mathcal{V}_m=0}
\end{equation}
is the transmission time used by Peres in \cite{Per80}. A similar expression can be obtained for the reflection time $t_c^R(k)$ in terms of the reflection phase $\varphi_R^{(m)}(k)$. It can be \emph{proved} that these \emph{clock times} satisfy \cite{CLM09} (also see \cite{SBa})
\begin{equation} \label{rel}
\tau_D(k) = |T(k)|^2t_c^T(k) + |R(k)|^2t_c^R(k),
\end{equation}
where $\tau_D(k)$ stands for the stationary dwell time \cite{But83}, and $T(k)$ and $R(k)$ are the transmission and reflection amplitudes, respectively. It is important to stress that, for the clock times used here, the above relation can be proved to be a \emph{direct consequence} of Schr\"{o}dinger's equation. Thus the criticism to such a relation for supposedly not being compatible with quantum mechanics due to the absence of an interference term, which sometimes appear in the literature (see, e.g., \cite{LMa94}), does not apply to the present case (for details see \cite{CLM09}).

It should be noticed that, in order to strictly enforce the conditions $|E|\gg\mathcal{V}_m$ and $|V(z)-E|\gg\mathcal{V}_m$, the integral in (\ref{psi_tr}) should in principle be truncated by a small cutoff excluding the regions in which these conditions are not warranted. However, the continuity of the integrand allow us to extend the integration over the whole $k$-space up to first order for a very small $\mathcal{V}_m$, consistent with the approximations in (\ref{1order}). This requires a large time resolution $\tau$ for the clock, since it is related to the reciprocal of $\mathcal{V}_m$ \cite{Lea93}. However, as observed by Davies \cite{Dav05}, there is no difficulty since the clock can be treated within the weak measurement theory of Aharonov \textit{et al.} \cite{AAV88, AVa90} and a precise value can be obtained for a large enough ensemble.

Here we are interested in finding an average \emph{transmission} time. Therefore, we must consider only the sub-ensemble of all scattered particles that are eventually transmitted, i.e., we must perform a \emph{post-selection} of the final state (in addition to the \emph{pre-selection} of the initial one). The post-selected (transmitted) state can be described, with a slight change in notation, by the density matrix of the particle-clock asymptotic state $\varrho^{\textrm{trans}} \equiv \alpha \left|\Psi_{trans}\right\rangle\left\langle \Psi_{trans}\right| $, where
$\alpha$ is a normalization constant and $\left|\Psi_{trans}\right\rangle$ corresponds to the state (\ref{psi_tr}). In order to observe the final state of the clock we assume an \emph{asymptotic condition}, stating that long after the interaction the transmitted wave packet is completely localized to the right of the potential. Then, we can trace out the particle's degrees of freedom (by extending the $z$-integral
over the whole space) and obtaining the clock's reduced density matrix for the transmitted sub-ensemble, given by
\begin{eqnarray}\nonumber
\!\!\!\!\!\!\! \varrho_{c}^{\textrm{trans}} &\!\!=\!\!& \alpha \int \frac{dk}{2\pi} \left|A(k)T(k)\right|^2 \\
&&\times \left|v_0\left(\theta-\omega t_c^T(k)\right)\right\rangle \left\langle v_0\left(\theta-\omega t_c^T(k)\right)\right|\, ,\label{dc}
\end{eqnarray}
where the normalization requirement ${\rm Tr} (\varrho_{c}^{\textrm{trans}}) =1$ implies $\alpha \!=\! \left\{\int \frac{dk}{2\pi} |A(k)T(k)|^2 \right\}^{-1}$.

Now, since the clock's reading is associated with the peak of its wave function, we \emph{define} the \textit{average transmission time} as $\langle t_c^T\rangle ={\rm Tr}\left( \varrho_{c}^{\textrm{trans}} \frac{\hat{\theta}}{\omega}\right)$,
where $\hat{\theta}$ is an operator giving the peak of $v_k$ \cite{AMM03}. Then,
\begin{equation}\label{time-av}
\langle t_c^T \rangle = \int dk  \, \rho_T(k) \, t_c^T(k)\, ,
\end{equation}
where $\rho_T(k)\!=\!\frac{\left| A(k)T(k)\right|^2}{\int dk \left| A(k)T(k)\right|^2 }$ is the probability density to find mode $k$ in the final transmitted wave packet. An analogous procedure for the reflected sub-ensemble results in the \textit{average reflection time}:
\begin{equation}\label{time-avR}
\langle t_c^R \rangle = \int dk \, \rho_R(k) \, t_c^R(k)\, ,
\end{equation}
with $\rho_R(k)\!=\!\frac{\left| A(k)R(k)\right|^2}{\int dk \left| A(k)R(k)\right|^2 }$ the corresponding probability density to find mode $k$ in the reflected wave packet.

In the special case of a very sharply peaked wave packet, $|A(k)|^2/(2\pi) \simeq \delta (k - k_0)$, definitions (\ref{time-av})-(\ref{time-avR}) reduce to the corresponding stationary times, as expected. However, for more physical incident wave packets, which necessarily involve a finite dispersion in wave numbers, these definitions have the important property of weighting the contribution of each $t_c^T(k)$ ($t_c^R(k)$) by the \emph{probability density of having the mode $k$ in the final transmitted (reflected) wave packet}, thus minimizing the contributions of modes leading to the most improbable results. This is in contradistinction with the approach of considering the density matrix of the transmitted and reflected wave packets together, which is equivalent to averaging the dwell time with respect to the initial wave packet components -- a procedure that leads to many well known difficulties \cite{Par09, Nus00}.

The expressions (\ref{time-av})-(\ref{time-avR}) coincide with the real part of the \emph{complex time} obtained by Leavens and Aers \cite{LAe89} in their analysis of double barrier resonant times using the Larmor clock (see also \cite{LAe89-2} which deals with similar averages for the time obtained in \cite{HFF87}). These definitions could also be inferred from the analysis of the Larmor clock by Falck and Hauge \cite{FHa88}. The main advantage of the above treatment (in which only the position of the clock's pointer is observed \cite{AVa90, Joz07}) is that it introduces a unique \textit{real} time scale, thus avoiding the problem of interpreting complex times.

The above average reflection and transmission times can be used to generalize relation (\ref{rel}) for a wave packet. First, we observe that the
asymptotic probability to find the particle in the region at the right of the potential (i.e., the total probability for the particle to be transmitted) is given by $P_T\!=\lim_{t\to\infty}\!\int_a^\infty dz \int_0^{2\pi}d\theta |\Psi_{trans}(\theta ,z,t)|^2$. Under the asymptotic condition stated above the
$z$-integration can be extended to the whole space and this probability reduces to
\begin{equation}\label{pt}
P_T \!=\! \int\frac{dk}{2\pi}|A(k)T(k)|^2\, .
\end{equation}
Analogously, assuming a similar asymptotic condition stating that long after the interaction the reflected wave packet is completely localized to the left of the potential, the total probability for the particle to be reflected is
$P_R =\int\frac{dk}{2\pi}|A(k)R(k)|^2$. From the normalization of the initial wave packet these probabilities satisfy $P_T+P_R\!=\!1$ (as required, since long after the interaction transmission and reflection are complementary outcomes). Then, defining the mean dwell time as \cite{Nus00, Par09}
\begin{equation} \label{dwell-av1}
\overline{\tau_D} \equiv \int \frac{dk}{2\pi}|A(k)|^2 \tau_D(k)\, ,
\end{equation}
where the overbar indicates averages over the \emph{initial} wave packet, the above
definitions, combined with (\ref{rel}), imply that
\begin{eqnarray}\nonumber
\overline{\tau_D} &\!\!=\!\!& \int \frac{dk}{2\pi}|A(k)T(k)|^2t_c^T(k) \\
&&+\, \int \frac{dk}{2\pi}|A(k)R(k)|^2t_c^R(k) \nonumber \\ \nonumber \\
&=& P_{T}\langle t_c^T \rangle +P_{R}\langle t_c^R \rangle\, . \label{dwell-av}
\end{eqnarray}
This relation says that the usually defined average of the dwell time over the initial wave packet (which does not take into account whether the particle is eventually reflected or transmitted) can be viewed as an average over the transmission and reflection average times, weighted by the total probabilities for transmission and reflection, respectively. It must be stressed that \emph{a necessary requirement} for this interpretation is that the \emph{asymptotic conditions be satisfied}. Relation (\ref{dwell-av}) appears as the real part of a similar \emph{complex} relation in \cite{LAe89} -- again, the advantage here is that the SWP clock provides only real times.

\section{Tunneling through symmetric barriers}
\label{tun}

In this section we study the average transmission time for a gaussian wave packet incident from the left on two kinds of symmetric localized potentials, namely a single rectangular barrier and the double delta barrier. In both cases, we shall be mainly concerned with the behavior of $\langle t_c^T \rangle$ in the regime of opaque barriers (characterized by very small transmission amplitudes for non resonant tunneling energies), in order to examine the emergence or not of the Hartman effect \cite{Har62} (single barrier) and its generalized version \cite{ORS02} (double delta barrier).

We assume that at $t=0$ the particle is in a state described by a right-moving gaussian wave packet centered around a tunneling wave number $k_0>0$, and
spatially centered around a position $z_0<0$, i.e.,
\begin{equation} \label{psi0}
\Phi (z, 0) = \frac{1}{(2\pi)^{1/4}\sqrt{\sigma}} \exp\left\{ ik_0 z - \frac{(z-z_0)^2}{4\sigma^2} \right\}\, ,
\end{equation}
where $\sigma$ is a scale measuring the spatial extent of the initial wave packet (its inverse is proportional to the wave packet's momentum dispersion).

In what follows we will always make use of the fact that, as a consequence of (\ref{rel}), for symmetric potentials $t_c^T(k)$ and $t_c^R(k)$ coincide with the dwell time $\tau_D(k)$ \cite{FHa88}. Therefore, the average times (\ref{time-av}) and (\ref{time-avR}) are just the dwell time averaged over the transmitted and reflected sub-ensembles, respectively.

\subsection{The single rectangular barrier}
\label{single}

Let us now consider a square potential barrier of height $V_0$ and width $a$, $V(z) = V_0\Theta (z) \Theta (a -z)$. For a tunneling incident energy, $E<V_0$, the
stationary dwell time is well known and given by \cite{But83}
\begin{eqnarray}\nonumber
   \!\!\!\! \tau_D(k)&\!=\!&\frac{2\mu}{\hbar}\frac{k}{q}\\
    &&\!\!\!\!\!\!\!\!\!\!\!\!\!\! \!\!\!\!\!\!\!\! \times\frac{\left[ (k^2+q^2)\tanh (qa) + qa(q^2-k^2){\rm sech}^2(qa) \right]}{\left[ 4q^2k^2 + (q^2-k^2)^2 \tanh^2(qa)\right]}\, ,\label{tc1B}
\end{eqnarray}
where $k=\frac{1}{\hbar}\sqrt{2\mu E}$ and $q=\frac{1}{\hbar}\sqrt{2\mu (V_0-E)}$. For energies above the barrier, $E>V_0$, the stationary dwell time can be obtained from the above expression through the substitution $q\to i k_1$, with $k_1=\frac{1}{\hbar}\sqrt{2 \mu (E-V_0)}$), while the transmission amplitude $T(k)$ can be found in any textbook on quantum mechanics (see, e.g., \cite{Mer}). In the opaque limit, characterized by $qa \gg 1$ for a given $E< V_0$, the stationary dwell time (\ref{tc1B}) becomes independent of the barrier width $a$, which is the Hartman effect \cite{Har62,ORS02}. This effect has been at the origin of an intense debate in the recent literature concerning, among other problems, the apparent superluminality of the tunneling process (see, for example, \cite{Win06-1,ORS02, Olk09,LMa07,CLM09} and references therein). It should be noticed that a time can be defined, based on an analogy with fluid mechanics \cite{HCP74}, such that in the \emph{stationary case} it reduces to the dwell time divided by the transmission coefficient and, therefore, does not saturate with the barrier width \cite{HCP74} (also see \cite{MAl90, SCP90, GIA04}); but a complete analysis of \emph{time dependent} tunneling using such time scale is yet to be achieved \cite{HCP74, Lea91}.

In Figure 1 we investigate the typical behavior of the average times (\ref{time-av}) and (\ref{time-avR}) with increasing barrier widths. For comparison we have also plotted the average reflection time (\ref{time-avR}), the mean dwell time (\ref{dwell-av1}), and the time spent by the incident packet (characterized by its peak wavelength $k_0$) in the region $[0,a]$ in the absence of the potential, i.e., $t_{free}= \frac{\mu a}{\hbar k_0}$.
The choice of parameters in these figures ensures that the initial wave packet (\ref{psi0}) is, to an excellent approximation, confined to the left of the barrier (at $t=0$ the probability to find the particle in the region $z>0$ is less than $10^{-15}$). Also, tunneling components are dominant in the Fourier decomposition of the initial wave packet.
We observe that in the extreme opaque regime (in this case characterized by very large barrier widths), while $\overline{\tau_D}$ saturates to a value independent of $a$,  $\left\langle t^T_c\right\rangle$  increases linearly with the barrier width. In the inset, showing in detail the region of small to moderate barrier widths, we observe that for thin barriers all the average times are very similar. The similarity between $\left\langle t^T_c\right\rangle$ and $\left\langle t^R_c\right\rangle$ [hence, from (\ref{dwell-av}), their similarity with $\overline{\tau}_D$] is expected, since for thin barriers the post-selected probability density distributions $\rho_R$ and $\rho_T$ are almost indistinguishable from the incident one, $\rho_{inc}$. As the barrier thickness increases $\langle t^R_c\rangle$
becomes smaller than the free time and soon saturates, as expected for a wave packet centered in a tunneling wave number, since one would anticipate that the wave function cannot penetrate beyond a certain barrier depth. The average transmission time, $\left\langle t^T_c\right\rangle$, also becomes smaller than the free time for intermediate values of the barrier width, indicating that the barrier speeds up the wave packet in the transmission channel; however, in the very opaque region its behavior suffers an abrupt change, growing suddenly until it attains a regime of linear growth -- as shown for large $a$'s in Fig. 1, in this regime $\left\langle t^T_c\right\rangle > t_{free}$ (The wave packet width exerts an important role in determining the values of $a$ for which this abrupt change occurs, with the barrier width for that change increasing with $\sigma$; for a similar conclusion using a different time average see \cite{BSM94}). Therefore, we observe that $\left\langle t^T_c\right\rangle$ \emph{never saturates}, as one would expect of a well defined average transmission time. Finally, as it is well known, the mean dwell time initially increases with $a$ and soon saturates (the Hartman effect); this is a consequence of the fact that in the opaque region $P_R\simeq 1$ and the reflection term dominates the r.h.s of Eq. (\ref{dwell-av}).
\begin{center}
\begin{figure}
\includegraphics[width=9.5cm,height=5.5cm]{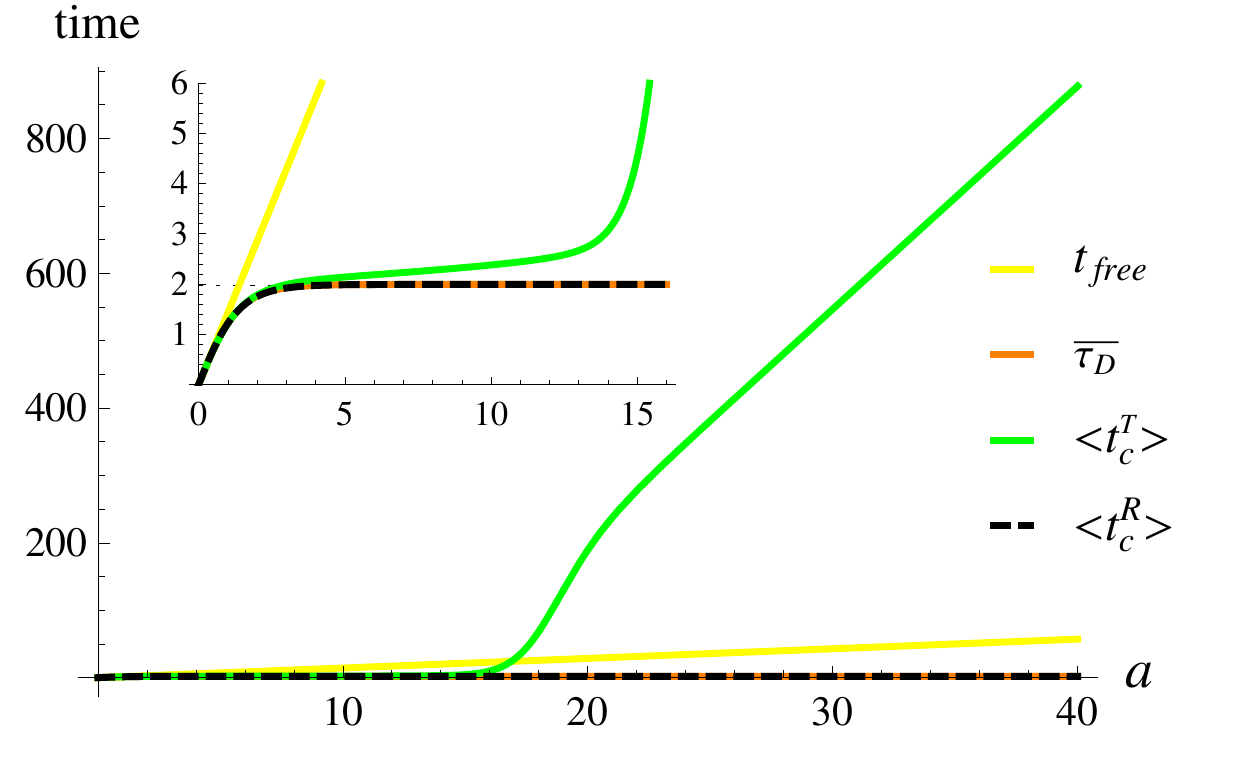}
\caption{(Color online) Behavior of the average times with respect to the barrier width $a$. All quantities are expressed in \emph{atomic units} (a.u.): $\mu=1$, $\sigma=10$, $k_0=0.7$, $z_0=-8\sigma$, and $V_0=0.5$. \emph{Inset}: \emph{zoom} of the region of small and moderate barrier widths [the horizontal dotted line indicates the saturated value of the stationary dwell time $\tau_D(k_0)$]. The free time, $t_{free}$, is plotted for comparison. \label{fig1}}
\end{figure}
\end{center}

It is instructive to compare the \emph{qualitative} behavior of the average times shown in Fig. \ref{fig1} with the results presented in \cite{GIA04}, even though \cite{GIA04} considers the time \emph{independent} situation. In both cases the reflection times saturate and the transmission times always increase with the barrier width, not showing any sign of the Hartman effect. However, there are important differences: in \cite{GIA04} the reflection time starts from infinite for $a=0$ and it is very large for very small barrier widths before staurating; furthermore, in \cite{GIA04} the transmission time tends to increase exponentially with $a$ in the extreme opaque limit (also see \cite{HCP74, MAl90, SCP90, Hag93}). These are unexpected properties and there is some debate about the interpretation of these times \cite{HCP74, Hag93, Win06-1} (nevertheless, the transmission dwell time used in \cite{GIA04} has been shown to have practical use in the study of the half life of radioactive nuclei decaying by the emission of an $\alpha$-particle \cite{KCN09}). By contrast, $\langle t^R_c\rangle$ is small for small barrier widths, a plausible result since this time is interpreted as the average time spent into the potential region by the wave packet which is eventually reflected. In addition, $\langle t^T_c\rangle$ tends to increase \emph{linearly} with $a$ in the extreme opaque limit. Such good properties of $\langle t^T_c\rangle$ and $\langle t^R_c\rangle$ emerge as a consequence of averaging over pos-selected sub-ensembles.

The behavior of the average transmission time observed in Fig. 1 is a consequence of the fact that for the transmitted wave packet in the opaque regime $\left\langle t_c^T \right\rangle$ is dominated by above the barrier components. If one truncates wave packet to include only under the barrier components, the corresponding ``truncated" transmitted wave packet is dominated by the contribution of a single component, completely lacking localizability in the opaque limit and approaching a plane wave and, therefore, saturating [but in this case, strictly speaking, the definition (\ref{time-av}) does not apply]. The above the barrier dominance is, of course, well known (by the way, methods to avoid such dominance, by compensating for the reflection of lower $k$ components in the state preparation, have been developed and result in average times that saturate \cite{HSM04}).

\subsection{The double delta barrier}
\label{double}

The \textit{generalized} Hartman effect states that for two potential barriers the tunneling time becomes independent of the distance between the barriers in the opaque limit \cite{ORS02, LLB02, Esp03, Win05} (see, however, \cite{LRo05} for an alternative view of the \emph{stationary} case using a multiple peak decomposition). In order to investigate this effect we will consider the double delta barrier $V(z) = \gamma\delta (z) + \gamma\delta (z-d)$, where $\gamma>0$ gives the barrier strength and $d$ is the separation between the barriers and we will assume that the SWP clock runs only when the particle is in the region $0<z<d$. This potential has the advantages of being the easiest way to study the generalized Hartman effect \cite{AER03} and that all the wave packet components are tunneling (for a study of stationary traversal times for this potential, see \cite{SBC94}).

For a stationary wave incident from the left, the transmission coefficient is given by (in the limit $\mathcal{V}_m = 0$)
\begin{equation}
\left|T(k)\right|^2=\frac{1}{1+4\alpha^2\left[\alpha\sin(kd)+\cos(kd)\right]^2}
\label{amptransdelta}.
\end{equation}
where $\alpha \equiv \frac{\mu\gamma}{\hbar^2 k}$. In the opaque limit, characterized by $\alpha \gg 1$ (equivalently, $\gamma \gg k$), $|T(k)|^2$ goes to zero almost everywhere (as $\gamma^{-4}$), except at the (countable) infinite set of resonances $k_n$. The resonances are given, for \emph{any} $\gamma>0$ and $d>0$, by the solutions of the transcendental equation \cite{Ricco}
\begin{equation}
k=-\frac{1}{d} \tan^{-1}\left(\frac{\hbar^2 k}{\mu\gamma}\right)\label{rescond},
\end{equation}
with the location of the $n$th resonance strongly dependent on $d$, in such way that the spacing between successive resonances decreases with $d$ (see \cite{POl05} and references therein). For $\gamma\to\infty$ the resonance spectrum tends to the spectrum of an infinite rectangular well of width $d$,  i.e. $k_n=\frac{n\pi}{d}$ (with $n=\pm 1,\pm 2,\pm 3,\cdots$) \cite{Mer}.

The stationary dwell time in the region between the barriers, as follows from (\ref{tc}), is given by
\begin{eqnarray}
&&\!\!\!\!\!\!\!\!\!\!\!\!\!\!\!\!\!\!\!\!\tau_D(k)= t_c(k) = \frac{\mu}{\hbar k^2}\nonumber\\
 &&\!\!\!\!\!\!\!\!\!\!\times \,\frac{\left( 1+2\alpha^2\right)kd+2\alpha\sin^2(kd)-\alpha^2 \sin(2kd)}{1+4\alpha^2\left[\alpha\sin(kd)+\cos(kd)\right]^2},\label{stattimedelta}
\end{eqnarray}
and at the $n$th resonance, $k_n$, its value is
\begin{equation}
\tau_D(k_n)=\frac{\mu}{\hbar k_n^3} \left[2\frac{\mu\gamma}{\hbar^2} +\left(2 \frac{\mu^2\gamma^2}{\hbar^4}+k_n^2\right)\, d\right],\label{restime}
\end{equation}
as can be found by substituting (\ref{rescond}) into  (\ref{stattimedelta}). For a fixed $d$, the opaque limit of the resonant dwell time behaves asymptotically as $\tau_D(k_n)\sim \frac{2 \mu^3}{\hbar^5}\frac{\gamma^2}{k_n^3}\, d$, thus diverging in the extreme opaque limit  $\gamma\to\infty$. On the other hand, the off resonance dwell time behaves in the opaque limit as $ \tau_D(k) \propto \gamma^{-2}$, thus characterizing the generalized Hartman effect by the fact that $\tau_D(k)\to 0$ as $\gamma\to\infty$ (however, such a convergence to zero is \emph{non-uniform} in any neighborhood of a resonance).
\begin{center}
\begin{figure}
\includegraphics[width=9.0cm,height=5.5cm]{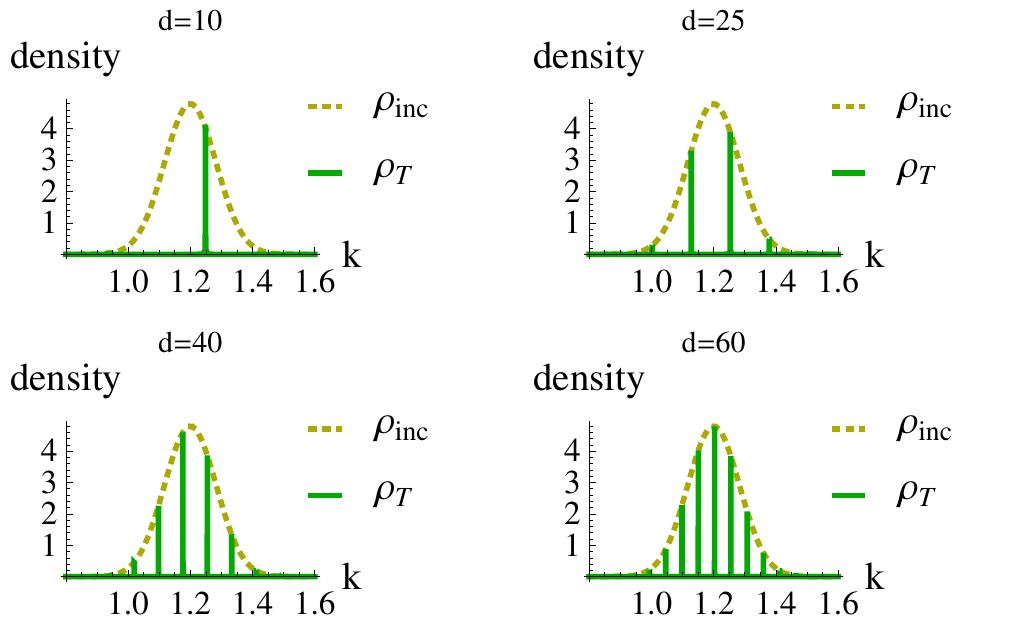}
\caption{(Color online) Wave number distribution for the transmitted (solid curve, \emph{non normalized}) and the incident wave packets (dashed curve), for opaque barriers. Atomic units are used, with $\gamma=16$, $\sigma=6$ and $k_0=1.2$. \label{fig4}}
\end{figure}
\end{center}

\begin{figure}
\includegraphics[width=8.5cm,height=5.0cm]{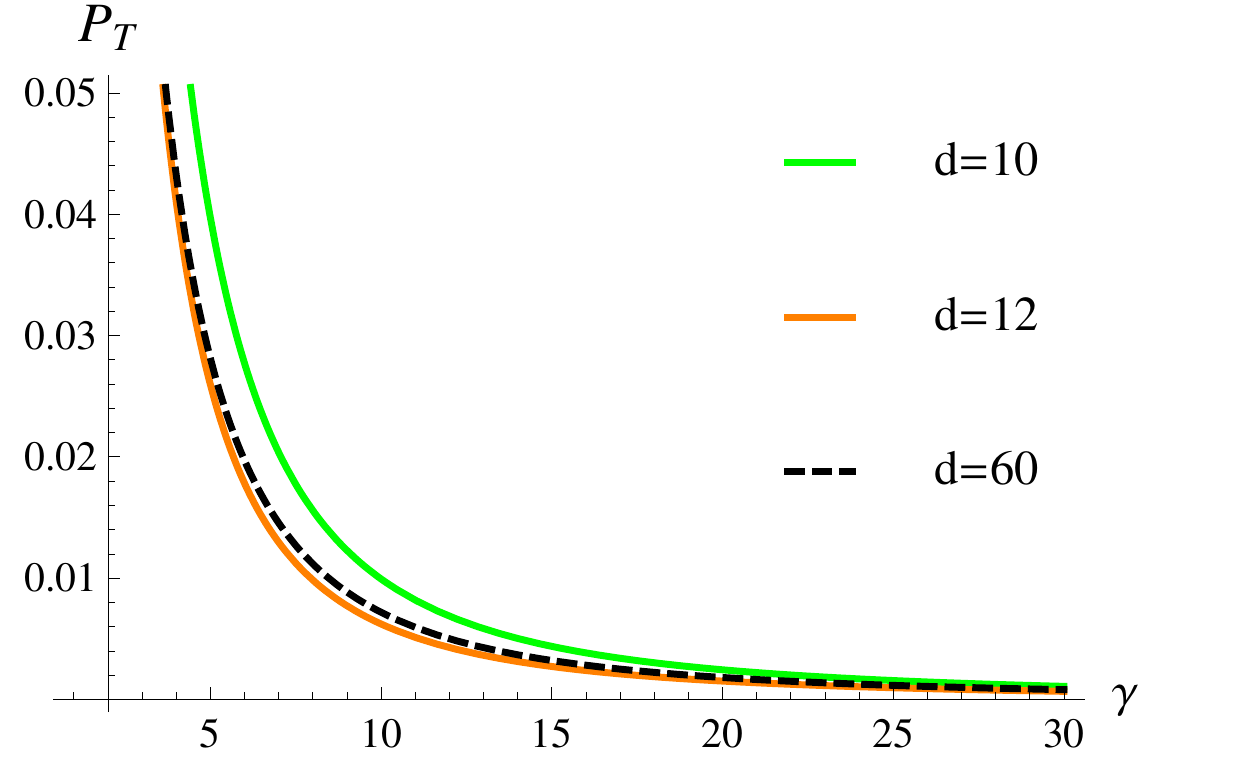}
\caption{(Color online)  Behavior of the total probability for the particle transmission with increasing strength $\gamma$. We observe that this probability goes to zero as the barrier strength $\gamma$ increases, irrespective the eventually large number of resonant components in the initial packet wave number distribution. In this plot $k_0=1.2$ and $\sigma=6$ in atomic units. \label{fig5}}
\end{figure}

Let us now consider a gaussian wave packet sharply peaked around $k_0$, eq. (\ref{psi0}), and investigate the behavior of the average transmission time in the opaque limit (which is now characterized by $\gamma \gg k_0$). It is useful to first analyze the
wave-number distribution of the transmitted and incident wave packets, given by the probability densities $\rho_T(k)$ and $\rho_{\textrm{inc}}(k)\equiv |A(k)|^2/(2\pi )$, respectively. As shown in Figure \ref{fig4}, for small $d$ there are few, if any, resonances in the domain of $\rho_{\mathrm{inc}}(k)$. However, as $d$ increases the resonances become densely distributed within that domain and we expect the transmitted packet to be dominated by them. Such a dominance is especially pronounced in the opaque regime, where the off resonance components are strongly suppressed (a similar behavior was found for the double \textit{square} barriers in \cite{POl05}). It is important to emphasize that even when the transmitted spectrum is dominated by resonances, \textit{in the extreme opaque limit the wave packet total transmission probability tends to vanish}, since in the limit $\gamma\to\infty$ the integrand in (\ref{pt}) tends to zero almost everywhere, except in a zero-measure set (the countable set of resonances) -- this is illustrated in Figure \ref{fig5}.

\begin{center}
\begin{figure}
\includegraphics[width=9.2cm,height=5.0cm]{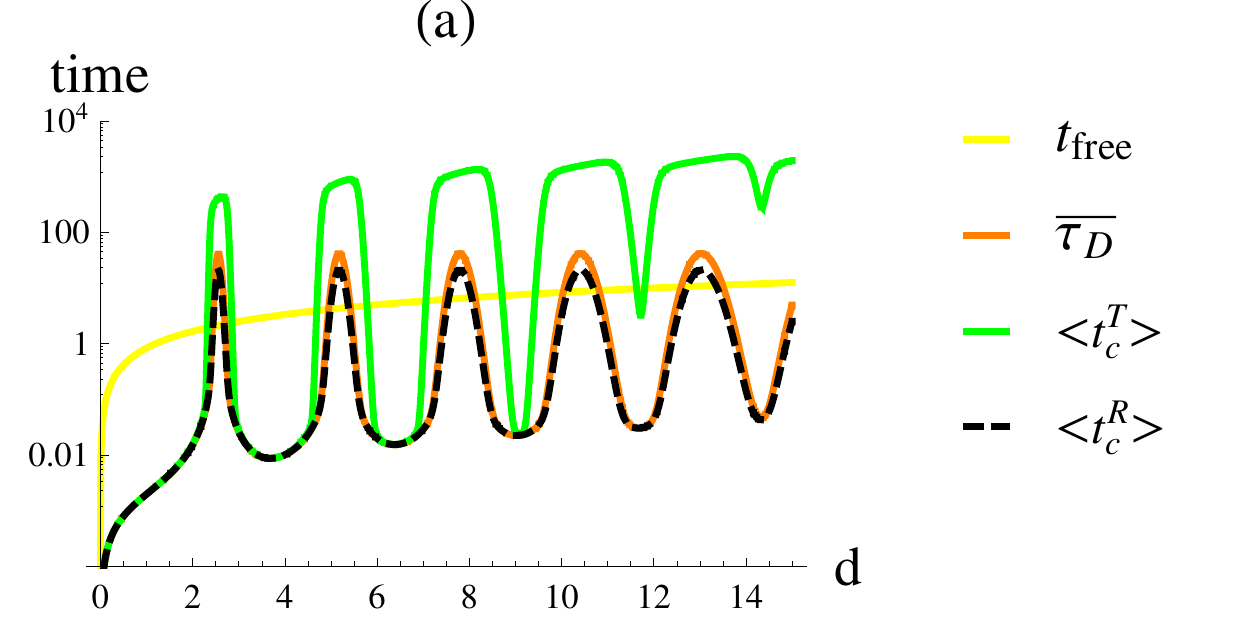}
\includegraphics[width=9.4cm,height=5.0cm]{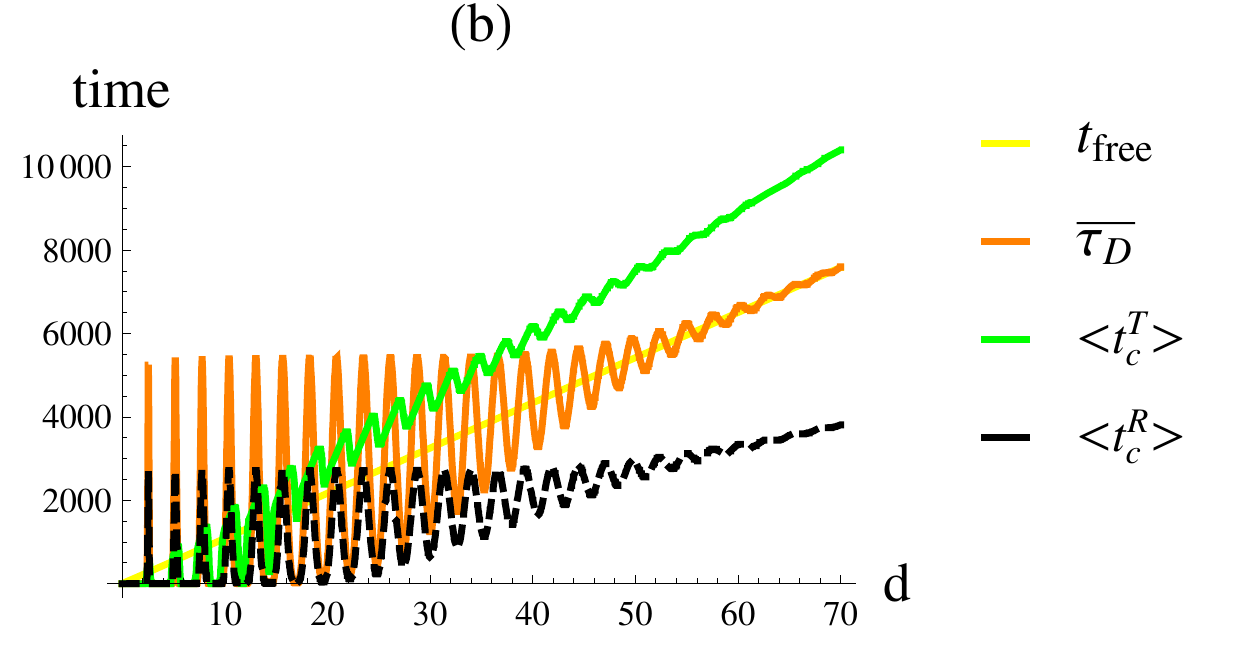}
\caption{(Color online) Behavior of the average transmission time $\left\langle t_c^T \right\rangle$ with respect to the barrier separation $d$ for a wave packet having a narrow wave number distribution around the central wave number $k_0$. All quantities are expressed in atomic units: $\gamma = 16$, $k_0=1.2$, $\sigma=20$, $z_0=-8\sigma$. The mean dwell time $\overline{\tau_D}$, the average reflection time $\left\langle t_c^R \right\rangle$, and the free time $t_{\mathrm{free}}$ are also shown for comparison. Fig (\emph{a}) corresponds to the range of small to moderate barrier widths (in which we used a log scale for the vertical axis). Fig. (\emph{b}) shows a wide range of barrier widths [we have multiplied all the times, except $\left\langle t_c^T \right\rangle$, by a convenient factor ($=130$), in order to display all the curves in the same plot]. \label{fig6}}
\end{figure}
\end{center}

Figure \ref{fig6}a displays $\left\langle t_c^R \right\rangle$, $\left\langle t_c^T \right\rangle$ and $\overline{\tau_D}$, in a logarithmic scale, for small to moderate barrier separations, and for $\gamma$ in the opaque domain. One observes that the average reflection time only deviates significantly from the mean dwell time when resonances are present around $k_0$, since there the reflection coefficient $|R|^2$ is very small. Thus, when resonances are densely distributed around $k_0$ (e.g., when $d$ is large) one expects significative differences between $\left\langle t_c^R \right\rangle$ and $\overline{\tau_D}$, as it is shown in Fig \ref{fig6}b, for vary large $d$'s.

In Figure \ref{fig6}b we observe the typical behavior of the average times for a wide range of values of the barrier separation $d$. Contrary to the generalized Hartman effect statement, this figure reveals that $\left\langle t_c^T \right\rangle$ always varies with $d$ (the plot corresponds to an initial wave packet having a narrow width $\Delta k\thicksim \frac{1}{\sigma}$ around $k_0$) and, in fact, for large enough barrier spacing $\left\langle t_c^T \right\rangle$ tends to grow linearly with $d$ (although not shown here, we observed a similar behavior for several sets of values of $\gamma$ and $\sigma$). This is a consequence of the fact that for large $d$ any small neighborhood of a given $k$ will be densely populated by resonances (see Fig. \ref{fig4}) and, as $d$ varies, these resonances (of different orders) enter and exit that neighborhood, with the net effect of a linear dependence on $d$ according to eq. (\ref{restime}) (with $n$ \emph{not} fixed, in a way that $k_n$ is always within such a neighborhood). On the other hand, the behavior of $\left\langle t_c^T \right\rangle$ for small and moderate barrier separations $d$ tends to be oscillatory because there are few resonances in the transmitted spectrum (see Fig. \ref{fig5}) and, as $d$ increases and a new single resonance eventually enters the domain of the transmitted spectrum, the peak of $\tau_D(k)$ at that resonance causes a peak in $\left\langle t_c^T \right\rangle$.

Summarizing, Figures \ref{fig6}a and \ref{fig6}b show that the average transmission time depends on $d$, showing no evidence of the generalized Hartman effect. This is in opposition to the conclusions of \cite{POl05,POl06} (using a different definition of average time, but with similar parameters) who assert that in regions of small $d$ located between the peaks associated to successive resonances (such that $ k_n - k_{n-1}\gg \Delta k $) the generalized Hartman effect occurs, as characterized by plateaus in between the peaks in the plot \emph{time} versus $d$ (here such plateaus would be at zero) -- no such plateaus appear for $\left\langle t_c^T \right\rangle$ (in fact, not even for $\overline{\tau_D}$) in the corresponding regions in Fig. \ref{fig6}a. In addition, Figure \ref{fig6}b shows that in the opaque regime $\gamma>>1$ the mean dwell time is dominated by the average reflection time, as one would expect from relation (\ref{dwell-av}), and for very large $d$  the mean dwell time tends to behave as the free time. In that range of $d$, $\left\langle t_c^T \right\rangle$ shows that the particle slows down for large $\gamma$, as expected due to the multiple reflections occurring in the region between the barriers.

%%%%%%%%%%%%%%%%%%%%%%%%%%
\section{Concluding Remarks}
\label{concl}

We considered the extension of the SWP clock to treat the one-dimensional tunneling of a particle, represented by a wave packet, through a static potential and addressed the question of how much time a particle that is eventually transmitted (reflected) spends in the potential region. Taking into account that to properly analyze this question one must consider only the sub-ensemble of the scattered particles that are finally transmitted (reflected), which corresponds to a post-selection of the final state, the SWP clock attributes a well-defined average time for these process, namely eqs. (\ref{time-av}) and (\ref{time-avR}). These are to be compared with the mean dwell time obtained in \cite{Par09} for the SWP clock. That $\langle
t_c^T\rangle$ and $\langle t_c^R\rangle$ are good candidates to transmission and reflection
times also follows from the fact that they naturally satisfy the generalization of the ``weighted average rule" \cite{LAe89-2} for the time dependent case [eq. (\ref{dwell-av})], provided the asymptotic conditions are satisfied. Also, the above approach has the advantage of providing these average values as the only time scale, in contradistinction with previous treatments of the Larmor clock in which they appear as the real part of a complex time \cite{LAe89,LAe89-2,HFF87}.

We analyzed the properties of the average transmission and reflection times for the scattering of a gaussian wave packet, centered on a tunneling wave number, by a square barrier and a double delta barrier and found that they would deviate from the mean dwell time only when the transmission or the reflection coefficients were very small [for symmetric potentials $t_c^T(k)$ and $t_c^R(k)$ equal the stationary dwell time and the average times (\ref{time-av}) and (\ref{time-avR}) are an average dwell time weighted by the transmission and reflection probability density, respectively]. Then, by carefully considering the opaque limit we found that $ \langle t_c^T\rangle$ shows no evidence of either the Hartman effect \cite{Har62} (single barrier) or its generalized version \cite{ORS02} (double delta barrier), and it also shows that the tunneling particle is slowed down
by extremely opaque barriers. These properties indicate that $ \langle t_c^T\rangle$ is a good candidate for an average transmission time. It should also be noticed that our conclusions concerning the emergence of the generalized Hartman effect are in disagreement with those of \cite{POl05, POl06} (who used a different definition for the average transmission time) \emph{in the region of small barrier spacing}, for they claim the existence of such effect for a wave packet as long as the width $\Delta k$ of the wave packet is much smaller than the spacing between resonances, while our results for $ \left\langle t_c^T\right\rangle$ show that the generalized Hartman effect does not emerge in \emph{any} region of the parameters. We stress that \emph{average} tunneling times are the most one can expect due to the lack of a well defined time operator and the consequent impossibility to obtain an eigenvalue for time.

The above analysis shows that the saturation of the transmission time, a feature associated with the Hartman effect, is a consequence of not taking into account the fact that the definition of a meaningful traversal time associated to a tunneling particle requires localizability, a requirement impossible to be satisfied in the stationary case. In addition, it is also a consequence of not taking the probability of transmission into account properly.

It should be noticed that even though the average transmission time does not saturate, for intermediate barrier widths $ \left\langle t_c^T\right\rangle$ increases very slowly with $a$ (single barrier) or $d$ (double delta), so that the transmitted wave packet ``speeds up" in comparison to the free wave packet (see Figs. \ref{fig1} and \ref{fig6}). Although this phenomenon is well-known \cite{LMa94}, it could be interpreted as a ``remnant" of the Hartman effect and it deserves further investigation (in the realm of relativistic quantum mechanics) due to the fact that it could allow for (non-causal) superluminal average speeds in this range of barrier widths.

Finally, given the sensible properties exhibited by the average times $\langle t_c^R\rangle$ and $\langle t_c^T\rangle$, it would be interesting to consider its possible application to the analysis of quantum tunneling in graphene, a field of much current interest and in which many of the issues discussed in this paper are of practical relevance (see, e.g., \cite{GGu09} and references therein), including the controversy around the Hartman effect \cite{WCL09, DDr10}.

%%%%%%%%%%%%%%%%%%%%%%%%%%%%%%%%%%%%%%%%%%%

\section{Acknowledgments}

This work was partially supported by NASA Minnesota Space Grant Consortium (L.A.M.) and Concordia College (A.T.N.).

%

%% The Appendices part is started with the command \appendix;
%% appendix sections are then done as normal sections
%% \appendix

%% \section{}
%% \label{}

\end{document}